\documentclass[twocolumn,showpacs,preprintnumbers,amsmath,amssymb,prl]{revtex4}

\usepackage{graphicx}
\usepackage{dcolumn}
\usepackage{bm}

\begin{document}

\author{Peter B. Weichman$^1$ and Ranjan Mukhopadhyay$^2$}

\affiliation{$^1$BAE Systems, Advanced Information Technologies, 6
New England Executive Park, Burlington, MA 01803
\\
$^2$Department of Physics, Clark University, Worcester, MA 01610}

\title{Revisiting the dynamical exponent equality $z=d$ for
the dirty boson problem}


\begin{abstract}

It is shown that previous arguments leading to the equality $z=d$
($d$ being the spatial dimensionality) for the dynamical exponent
describing the Bose glass to superfluid transition may break down,
as apparently seen in recent simulations (Ref.\ \cite{Baranger}).
The key observation is that the major contribution to the
compressibility, which remains finite through the transition and was
predicted to scale as $\kappa \sim |\delta|^{(d-z)\nu}$ (where
$\delta$ is the deviation from criticality and $\nu$ is the
correlation length exponent) comes from the analytic part, not the
singular part of the free energy, and therefore is not restricted by
any conventional scaling hypothesis.

\end{abstract}

\pacs{
64.60.Fr,   
67.40.-w,   
72.15.Rn,   
74.78.-w,   
}

\maketitle

Beginning with the realization that the zero temperature onset
transition of superfluidity in a random medium should properly be
treated as a fluctuation driven quantum phase transition
\cite{MHL,FF,FWGF,WK}, a number of scaling arguments have been put
forward \cite{FF,FWGF} to determine or to place bounds on the
critical exponents describing it.  Probably the most significant of
these was an argument that the dynamical exponent, $z$, describing
the relative divergence of the temporal and spatial correlation
lengths via $\xi_\tau \sim \xi^z$, should be equal to the dimension,
$d$, of space.  The argument was supported by exact calculations in
$d = 1$, where indeed $z=1$ \cite{FWGF}, by $1 + \epsilon$
renormalization group calculations \cite{herbut} which, however, are
not completely rigorous due to the absence of a form for the
Hamiltonian for noninteger $d > 1$, and by a series of quantum Monte
Carlo (QMC) studies in $d \leq 2$ \cite{QMC,QMC2,Baranger}. Results
in $d > 2$ are restricted to a badly controlled double
$\epsilon$-expansion which is inappropriate for testing an exact
scaling argument of this type \cite{WK,MW}, but nevertheless points
to values of $z$ significantly larger than unity \cite{MW}.

With one controversial exception \cite{QMC2}, the earlier QMC
results \cite{QMC} were fit to $z=d$, but the numerical error bars
were not very tight. The more recent QMC study in $d = 2$
\cite{Baranger} provided a much more stringent test: using larger
systems and a joint, optimal critical functional form fit to three
separate thermodynamic quantities, the value $z = 1.40 \pm 0.02$ was
found, strongly violating $z=d$.  In light of this, the purpose of
this paper is to revisit the scaling arguments \cite{FF,FWGF} that
led to $z=d$, and show that they can in fact \emph{break down:} $z$
might be independent of all the other exponents. In the absence of
further theoretical justification, we raise the possibility that $z$
is unconstrained by any simple scaling argument.

Before presenting details, we briefly review the essence of the
scaling arguments and indicate where they could go wrong.  The
helicity modulus, $\Upsilon$ [or superfluid density, $\rho_s =
(m^2/\hbar^2) \Upsilon$], quantifies the response of the superfluid
to gradients in the phase, $\phi$, of the order parameter.
Specifically, the free energy density contains a correction $\Delta
f_x = (\Upsilon/2\beta V) \int_0^\beta d\tau \int d^dx |\nabla
\phi({\bf x},\tau)|^2$, where $V$ is the volume, $\beta=1/k_BT$, and
$\tau$ is the usual imaginary time variable. The essence of the
Josephson scaling argument is that since $\nabla \phi$ has
dimensions of inverse length, it should scale as $\xi^{-1} \sim
|\delta|^\nu$ where $\delta$ is the deviation of the control
parameter from criticality and $\nu$ is the correlation length
exponent.  On the other hand the singular part of the free energy,
$f_s$, is defined to scale as $|\delta|^{2-\alpha}$.  This implies
that $\Upsilon \sim |\delta|^\upsilon$ with $\upsilon = 2 -\alpha
-2\nu = (d+z-2)\nu$, where the last equality follows from the
quantum hyperscaling relation $2-\alpha = (d+z)\nu$ \cite{FWGF}.
Now, the Josephson relation, which connects changes in the chemical
potential to the time derivative of the phase, actually allows one
to interpret the \emph{compressibility}, $\kappa$, as a helicity
modulus in the imaginary time direction.  Thus $\kappa$ enters a
free energy correction $\Delta f_\tau = (\kappa/2\beta V)
\int_0^\beta d\tau \int d^dx (\partial_\tau \phi)^2$.  The same
logic thenn suggests that $\partial_\tau \phi$ should scale as
$\xi_\tau^{-1} \sim |\delta|^{z\nu}$, leading to $\kappa \sim
|\delta|^{\upsilon_\tau}$ with $\upsilon_\tau = 2-\alpha-2z\nu =
(d-z)\nu$.  Since both the superfluid and the Bose glass phases have
finite compressibility, one expects on physical grounds that
$\kappa$ should be finite and nonzero at $\delta = 0$ as well. This
immediately leads to the prediction $z=d$ \cite{FF,FWGF}. However, a
hidden assumption in these arguments is that $\Delta f_x$ and
$\Delta f_\tau$ can be included in the singular part of the free
energy, $f_s$.  We shall show that this is correct for $\Delta f_x$
but not for $\Delta f_\tau$. In fact the main contribution to
$\Delta f_\tau$ comes from the nonsingular analytic part of the free
energy, $f_a$, so that $\kappa$ is dominated by its analytic part
which is trivially finite at the transition, and $f_s$ yields only
small corrections to this that vanish at the critical point.

Rigorous definitions of the helicity modulii involve comparing free
energies with twisted and untwisted boundary conditions. Previously
such boundary condition dependence, $\Delta f$, has always been
included in $f_s$:  normally the strong dependence required for a
finite helicity modulus requires long or quasi-long range order that
is present only in the superfluid phase.  The fact that $\Upsilon
\equiv 0$ in the disordered phase then guarantees that it can arise
only from singular terms in the free energy.  In the present
problem, however, $\kappa$ is nonzero in both phases: the gapless
excitation spectrum of the Bose glass phase leads to power law order
in imaginary time (though not in space) \cite{FWGF}. The free energy
of both phases therefore have strong temporal boundary condition
dependence, and an analytic contribution is actually very natural.

From our analysis there emerges the following general criterion for
when $\Delta f$ can be dominated by $f_a$.  If the twisted boundary
condition breaks a fundamental symmetry of the model, then it can be
expected to generate relevant (in the renormalization group sense)
terms in the Lagrangian.  This leads to a new relevant, diverging
scaling variable in the singular part of the free energy that
generally will dominate all other contributions to $\Delta f$.  In
this case the usual Josephson scaling relation will hold.  On the
other hand, if no additional symmetry is broken, no new relevant
scaling variable results and the twist will lead only to small
shifts in the parameters already present in the untwisted
Lagrangian.  The helicity modulus, which involves derivatives with
respect to these shifts, will then be dominated by $f_a$.  In the
present problem the twist couples to particle-hole symmetry
\cite{MW} which is always broken at the Bose glass to superfluid
transition.

For convenience we consider a continuum $|\psi|^4$-model functional
integral representation of the partition function, $Z = \int D\psi
\exp(-{\cal L}_B)$, with Lagrangian \cite{FWGF,MW}
\begin{eqnarray}
{\cal L}_B &=& \int_0^\beta d\tau \int d^dx
\bigg\{-J \psi^*({\bf x},\tau) \nabla^2 \psi({\bf x},\tau)
\nonumber \\
&&-\ K \psi^*({\bf x},\tau) [\partial_\tau - \mu({\bf x})]^2
\psi({\bf x},\tau)
\nonumber \\
&&+\ r({\bf x})|\psi({\bf x},\tau)|^2
+ u|\psi({\bf x},\tau)|^4 \bigg\},
\label{Joe}
\end{eqnarray}
where $J \approx \hbar^2/2m > 0$ is the boson hopping amplitude
which favors spatial ferromagnetic order in the phase of the field
$\psi$; $\mu({\bf x})$ is the (static) random external potential;
and $K \approx 1/2u_0 > 0$, where $u_0$ is the soft core repulsion
between nearby bosons, favors ferromagnetic order along the
imaginary time dimension as well.  This model is a continuum
approximation to the Josephson junction array Lagrangian in which
$\psi({\bf x},\tau) \to \exp[i\phi_i(\tau)]$ where $\phi_i(\tau)$ is
the Josephson phase at site $i$, and the $\psi^* \nabla^2 \psi$ term
is the continuum limit of the Josephson coupling, $J_{ij}
\cos[\phi_i(\tau) - \phi_j(\tau)]$, between two nearby sites $i$ and
$j$.  The $r|\psi|^2 + u|\psi|^4$ terms then represent the usual
soft spin approximation to the the constraint $|\psi|=1$. We allow
$r({\bf x})$ to be random as well, representing disorder in the
hopping amplitudes, $J_{ij}$.  We could also allow $J$ and $u$ to be
random, but this is less convenient and does not produce anything
new.  We write $\mu({\bf x}) = \mu - \epsilon({\bf x})$ and $r({\bf
x}) = r_0 + w({\bf r})$ with disorder average $[\epsilon({\bf
x})]_{av} = [w({\bf x})]_{av} = 0$. For a given value of $\mu$, the
transition to superfluidity occurs with increasing $J$ at a point
$J_c(\mu)$. and we shall take $\delta = J - J_c(\mu)$.

Other representations of the problem serve equally well.  The
arguments we shall present are very general and do not depend on the
precise form of ${\cal L}_B$.  As seen below, the only common
feature required is that time derivatives and chemical potential
always appear in the combination $(\partial_\tau - \mu)$. In the
coherent state functional integral formulation, for example, the
Lagrangian has only a linear term, $\psi^* (\partial_\tau -
\mu)\psi$.  The two formulations match up to irrelevant terms (in
the renormalization group sense) so long as $\mu({\bf x})$ does not
vanish identically \cite{FWGF,MW}.  When $\mu({\bf x}) \equiv 0$ the
model (\ref{Joe}) reduces to the well known random rod problem,
corresponding to a classical $(d+1)$-dimensional $XY$-model with
columnar disorder \cite{MW,DBC,WM}. This model maintains an exact
particle-hole symmetry, and the Josephson scaling relation for
$\kappa$ is therefore valid, but with $z < d$: see below.

Twisted \emph{$\theta$-boundary conditions} \cite{FBJ} are defined
by the condition
\begin{equation}
\psi(x+L_\alpha{\bf \hat e}_\alpha) = e^{i\theta_\alpha} \psi(x),
\ \ |\theta_\alpha| \leq \pi,
\label{theta}
\end{equation}
where $x \equiv (\tau,{\bf x})$, ${\bf \hat e}_\alpha$, $\alpha=0,1,
\ldots,d$, are the space-time unit vectors, and $L_\alpha$ is the
dimension of the system along ${\bf \hat e}_\alpha$ (with
$L_0=\beta$). Thus $\bar \psi({\bf x}) = e^{-i{\bf k}_0 \cdot {\bf
x}-i\omega_0\tau} \psi({\bf x})$, where $\omega_0 = \theta_0/\beta$
and ${\bf k}_0 = (\theta_1/L_1,\ldots,\theta_d/L_d)$, obeys periodic
boundary conditions.  We therefore expect $\bar \psi_0 = \langle
\bar \psi({\bf x}) \rangle$ to be uniform and hence the order
parameter $\langle \psi({\bf x}) \rangle = e^{i{\bf k}_0 \cdot {\bf
x} + i\omega_0\tau} \bar \psi_0$ to have a uniform phase twist.  We
now rewrite the Lagrangian in terms of $\bar \psi$, obtaining
\begin{equation}
{\cal L}_B^{\omega_0 {\bf k}_0}[\psi;\mu,r_0]
= {\cal L}_B[\bar \psi;\mu - i\omega_0;r_0 + J {\bf k}_0^2]
+ {\bf k}_0 \cdot {\bf P}[\bar\psi].
\label{Ltwist}
\end{equation}
where ${\bf P}[\bar\psi] = -iJ \int d^dx \int_0^\beta d\tau [\bar
\psi^* \nabla \bar \psi - \bar \psi \nabla \bar \psi^*]$ is the
total momentum. The simple shift, $\mu \to \mu-i\omega_0$, is
guaranteed by the combination $(\partial_\tau - \mu)$ for any
model of dirty bosons.

The free energy is defined by $f^{{\bf k}_0 \omega_0} = (\beta
V)^{-1} [\ln(Z)]_{av}$, where the superscripts label the boundary
condition. Using (\ref{Ltwist}), straightforward expansion of the
free energy in powers of ${\bf k}_0$ at $\omega_0=0$,
\begin{equation}
\Delta f_x \equiv f^{{\bf k}_0} - f
= \frac{1}{2} \Upsilon {\bf k}_0^2 + O({\bf k}_0^4).
\label{delfx}
\end{equation}
produces the usual momentum-momentum type correlation function
expression for $\Upsilon$.  On the other hand, if ${\bf k}_0 = 0$ we
have the obvious result $f^{\omega_0}(\mu) = f(\mu-i\omega_0)$.
Analyticity of $f$ within a given thermodynamic phase immediately
implies the expansion
\begin{equation}
\Delta f_\tau \equiv f^{\omega_0} - f = i\omega_0 \rho
+ \frac{1}{2} \kappa \omega_0^2 + O(\omega_0^3),
\label{delft}
\end{equation}
where $\rho = -\partial f/\partial \mu$ and $\kappa = \partial \rho/
\partial \mu$.  It is the comparison of equations (\ref{delfx})
and (\ref{delft}) that motivates the identification of $\kappa$ with
the temporal helicity modulus.  Note, however, that the density
$\rho$ actually yields the leading term linear in $\omega_0$. The
reason this term appears in (\ref{delft}) but not in (\ref{delfx})
is that ${\cal L}_B$ contains only quadratic spatial derivatives and
is therefore even under space inversion, ${\bf x} \to -{\bf x}$.  On
the other hand ${\bf P}[\psi]$ is odd under space inversion,
implying that $f^{{\bf k}_0 \omega_0}$ is an even function of ${\bf
k}_0$.  Moreover we see that ${\bf k}_0$ \emph{breaks} the inversion
symmetry of ${\cal L}_B[\psi]$.  In contrast, only if $\mu({\bf x})
\equiv 0$ does ${\cal L}_B$ possess inversion symmetry in time,
$\tau \to -\tau$, which we call particle-hole symmetry \cite{foot1}.
This symmetry is already broken by $\mu$, and the additional twist,
$\omega_0$, produces nothing new.  These observations will be
crucial to understanding the scaling of $\Upsilon$ and $\kappa$.

Let us consider first the Josephson scaling argument for the random
rod problem, which then possesses both inversion and particle-hole
symmetry.  Classical intuition suggests in addition that $\kappa
\equiv 0$ in the disordered phase \cite{foot2}.  As alluded to
earlier, we then expect the leading boundary condition dependence to
enter via singular scaling combinations $k_0 \xi$ and $\omega_0
\xi_\tau$.  Formally, this expectation is implemented via a
finite-size scaling ansatz for the free energy density
\cite{FBJ,FWGF},
\begin{equation}
\Delta f^{{\bf k}_0 \omega_0} \approx \beta^{-1}L^{-d}
\Phi_0^{{\bf k}_0 \omega_0}(A_0 \delta L^{1/\nu_0},B_0
\delta \beta^{1/z_0\nu_0}),
\label{fscale}
\end{equation}
where $A_0$ and $B_0$ are nonuniversal scale factors \cite{KW}. For
convenience we take a hypercubical volume $V=L^d$, and the exponents
are those of the classical random rod problem.  The existence of a
nonzero stiffness, i.e., a leading finite-size correction of order
$L^{-2}$ or $\beta^{-2}$, now requires that the scaling function
$\Phi_0^{{\bf k}_0 \omega_0}(x,y) \approx x^{d\nu_0} y^{z_0\nu_0}
(Q_x x^{-2\nu_0} + Q_\tau y^{-2z_0\nu_0})$ for large $x,y > 0$,
yielding
\begin{eqnarray}
\Upsilon &\approx& A_0^{(d-2)\nu_0} B_0^{z_0\nu_0}
(Q_x/\theta_x^2) \delta^{\upsilon_0}
\nonumber \\
\kappa &\approx& A_0^{d\nu_0} B_0^{-z_0\nu_0} (Q_0/\theta_0^2)
\delta^{\upsilon_{\tau 0}},
\end{eqnarray}
where $\theta_x^2 = \theta_1^2 + \ldots + \theta_d^2$, implying
the Josephson scaling relations $\upsilon_0 = (d+z_0-2) \nu_0 =
2-\alpha_0-2\nu_0$ and $\upsilon_{\tau 0}= (d-z_0)\nu_0 =
2-\alpha_0-2z_0\nu_0$, and requiring in addition $Q_{x,0} \propto
\theta_{x,0}^2$.

Consider now the Bose glass to superfluid transition.  The density
and compressibility of both phases are smooth nontrivial functions
of $\mu$, and a temporal twist now perturbs only slightly the
particle-hole symmetry breaking term that is already in ${\cal
L}_B$. On the other hand, spatial twists still produce singular
corrections, and \emph{at} $\omega_0=0$ we predict a finite-size
scaling form \cite{foot5}
\begin{equation}
\Delta f^{{\bf k}_0} = \beta^{-1} L^{-d}
\Phi^{{\bf k}_0}(A \delta L^{1/\nu}, B \delta \beta^{1/z\nu}),
\label{bscale}
\end{equation}
with $\Phi^{{\bf k}_0}(x,y) \approx R_x x^{(d-2)\nu} y^{z\nu}$ for
large $x,y>0$, yielding $\Upsilon \approx A^{(d-2)\nu} B^{z\nu}
(R_x/\theta_x^2) \delta^{\upsilon}$, $\upsilon = (d+z-2)\nu
=2-\alpha-2\nu$, and $R_x \propto \theta_x^2$ as before.  All
exponents now refer to the dirty boson critical point.  Because we
have, as yet, imposed no temporal twist, \emph{there can be no
$O(\beta^{-1},\beta^{-2})$ terms}.

Now, if a finite $\omega_0$ is included, the only changes in
(\ref{bscale}) are that $\mu \to \mu - i\omega_0$ everywhere, and
boundary condition dependence of $f_a$ must be included.  Thus
\begin{eqnarray}
\Delta f^{{\bf k}_0 \omega_0} &=& \beta^{-1} L^{-d}
\Phi^{{\bf k}_0}(A \delta_\theta L^{1/\nu},B \delta_\theta
\beta^{1/z\nu})
\label{bscale2} \\
&&+\ f_a(J,r_0+J{\bf k}_0^2,\mu-i\omega_0) - f_a(J,r_0,\mu),
\nonumber
\end{eqnarray}
where $\delta_\theta = J-J_c(\mu-i\omega_0) \approx \delta +
i\omega_0 J_c^\prime(\mu)$.  Most importantly, $\Phi^{{\bf k}_0}$
\emph{is the same function} as that in (\ref{bscale}), and therefore
still yields no $O(\beta^{-1},\beta^{-2})$ terms. This implies
immediately that the scaling function $\Phi^{{\bf k}_0 \omega_0}$
produces no \emph{direct} contributions to $\rho$ and $\kappa$,
which must therefore arise (a) \emph{indirectly} from the $\omega_0$
dependence of $\delta_\theta$, and (b) directly from the analytic
part of the free energy.  The former couples derivatives with
respect to $\mu$, or equivalently $\omega_0$, to derivatives with
respect to $\delta$, producing the leading singular terms, $\kappa_s
\sim |\delta|^{-\alpha}$ \cite{foot6}. However, $\alpha = 2 -
(d+z)\nu$ is very likely negative \cite{FWGF}, so this gives a
vanishing contribution at the critical point.  Therefore the main
contribution is analytic in origin.  Taking ${\bf k}_0 = 0$, we may
write $f_a(J,\mu) = -\rho_c(J) [\mu - \mu_c(J)] - \frac{1}{2}
\kappa_c(J)[\mu - \mu_c(J)]^2 + \ldots$, expanded for convenience
about the transition line $\mu_c(J)$, and we immediately obtain from
(\ref{bscale2}) a finite compressibility right through the
transition, with the exponent $z$ nevertheless \emph{undetermined}
\cite{foot3,foot7}.

There is a second approach to the derivation of the Josephson
scaling relation, through scaling of the two point correlation
function in the superfluid hydrodynamic regime, which we now
discuss.  Long wavelength, low frequency fluctuations are known to
be governed by the effective Gaussian Lagrangian \cite{FF,FWGF}
\begin{equation}
{\cal L}_\mathrm{HD} =
\frac{1}{2} \int d^dx \int_0^\beta d\tau
[\Upsilon |\nabla \tilde\phi|^2
+ \kappa (\partial_\tau \tilde\phi)^2],
\end{equation}
where $\tilde\phi({\bf x},\tau)$ is the coarse-grained phase,
related to the coarse grained order parameter field via
$\tilde\psi({\bf x}, \tau) = \psi_0 \exp[i\tilde\phi({\bf
x},\tau)]$, where the bulk order parameter, $\psi_0 \sim
|\delta|^\beta$, near criticality.  The small ${\bf k},\omega$ form
of the Fourier transform of the two point correlator, $G({\bf
x}-{\bf x}^\prime,\tau-\tau^\prime) = [\langle \psi^*({\bf x},\tau)
\psi({\bf x}^\prime,\tau^\prime) \rangle]_{av}$, is then
\begin{equation}
G({\bf k},\omega) \approx |\psi_0|^2/[\Upsilon {\bf k}^2
+ \kappa \omega^2].
\label{Ghydro}
\end{equation}
Normally it is assumed that $G$ obeys the scaling form
\begin{equation}
G({\bf k},\omega) \approx C|\delta|^{-\gamma}
g(D k |\delta|^{-\nu},E \omega |\delta|^{-z\nu}),
\label{Gscale}
\end{equation}
where $\gamma$ is the susceptibility exponent and $C$, $D$, $E$ are
again nonuniversal scale factors.  If one naively matches
(\ref{Ghydro}) and (\ref{Gscale}) one concludes that $g(x,y) \approx
(g_1 x^2 + g_2 y^2)^{-1}$ for small $x,y$, and hence that
\begin{eqnarray}
\Upsilon/|\psi_0|^2 &\approx& g_1 C^{-1}
D^2 |\delta|^{\gamma - 2\nu}
\nonumber \\
\kappa/|\psi_0|^2 &\approx& g_2 C^{-1} E^2
|\delta|^{\gamma - 2z\nu}.
\end{eqnarray}
Using the scaling relation $\alpha + 2\beta + \gamma = 2$ we obtain
again the Josephson relations $\upsilon = 2-\alpha-2\nu$,
$\upsilon_\tau = 2-\alpha-2z\nu$.

This argument, however, is just a disguised version of the free
energy argument: ${\cal L}_\mathrm{HD}$ is obtained by assuming that
the energetics describing global phase twists also describes slowly
varying local phase twists.  Thus, locally we replace ${\bf k}_0$ by
$\nabla \tilde\phi$ and $\omega_0$ by $\partial_\tau \tilde\phi$,
then integrate over all of space-time \cite{foot4}. Since $\kappa$
arises from the nonscaling part of the free energy, it is unlikely
that it can now arise from the scaling part of the two-point
function.  Rather, we must carefully reconsider the scaling ansatz
for $G$.  Our proposal (for which we have no detailed theoretical
support at this stage), is that
\begin{eqnarray}
|\psi_0|^2/G({\bf k},\omega) &\approx& H |\delta|^{2-\alpha}
\Gamma(D k |\delta|^{-\nu},E \omega |\delta|^{-z\nu})
\nonumber \\
&&+\ \Gamma_a({\bf k},\omega),
\end{eqnarray}
where $\Gamma_a$ is analytic.  This self-energy scaling form is very
similar to (\ref{bscale2}).  Matching with (\ref{Ghydro}), we again
assume that $\Upsilon$ arises from the scaling part, $\kappa$ from
the analytic part.  Thus $\Gamma(x,y) \approx \gamma_1 x^2 +
\gamma_2 y^2$, for small $x,y$ while $\Gamma_a(k,\omega) \approx
\kappa \omega^2$ for small $k,\omega$. The $x^2$ term yields
$\Upsilon \sim |\delta|^\upsilon$. If $z < d$ the $y^2$ term is
subdominant to the analytic term, which yields $\kappa$ finite, as
required, leaving $z$ undetermined \cite{nu_vs_dover2,Dohm}. If we
consider only static correlations, $\omega = 0$, standard scaling is
recovered without any unusual analytic corrections.

To summarize, the original scaling ansatz for the Bose
glass-superfluid critical point, in which temporal twists produce a
relevant symmetry breaking perturbation to the Lagrangian, and
therefore scale in the combination $\omega_0 \xi_\tau$, is not
supported theoretically, and may explain violations of $z = d$ in $d
= 2$ \cite{Baranger}.

\emph{Acknowledgments:} We thank D. S. Fisher for fruitful
discussions.


\begin{thebibliography}{99}
\bibitem{foot1} This must be contrasted with time reversal symmetry,
which involves a simultaneous replacement $\psi \leftrightarrow
\psi^*$, which ${\cal L}_B$ always satisfies.
\bibitem{foot6} Within the $\epsilon,\epsilon_\tau$-expansion
\cite{DBC}, where $\epsilon_\tau$ is the dimension of time and
$\epsilon = 4-d-\epsilon_\tau$, it was found \cite{MW} that the Bose
glass phase is compressible \emph{only} in the physical dimension,
$\epsilon_\tau = 1$, with $\kappa \equiv 0$ for any $\epsilon_\tau <
1$.  This precludes any analytic contribution to $\kappa$.  We
expect then that $\kappa \sim
|\delta|^{-\alpha(\epsilon,\epsilon_\tau)} \to 0$ from the
superfluid side.
\bibitem{foot7} This can be confirmed explicitly in
$d=1$ \cite{foot2}.  Except for rapidly decaying irrelevant terms,
$\mu$ can be removed entirely from the dual surface roughening model
Lagrangian via a shift, appearing, consistent with our general
arguments, only as an additive analytic term in the free energy. The
result $z=1$ for this model is therefore unconnected to the
appearance of a finite $\kappa$.
\bibitem{foot4}  The linear term, $i\rho (\partial_\tau \tilde
\phi)$ integrates to an integer multiple of $2\pi i \rho$ and
disappears from $\exp(-{\cal L}_\mathrm{HD})$ since the total boson
number $\rho V$ is an integer.
\end{thebibliography}
\end{document}